\documentclass{llncs}

\usepackage{times}
\usepackage{epsfig}
\usepackage{ifpdf}

\usepackage{array}
\usepackage{amsmath,amssymb,amscd,amstext}
\usepackage{mathrsfs}
\usepackage{dsfont}
\usepackage{amsbsy}
\usepackage{stmaryrd}

\usepackage{subfigure}

\usepackage{setspace}
\usepackage{fancyhdr} 
\usepackage{hyperref}
\usepackage{lastpage}
\usepackage{pslatex}

\usepackage[displaymath,textmath,sections,graphics,floats,graphicx,color]{preview}

\def  \ft#1     {\footnote{#1} }

\newcommand{\lts}[1]{\ensuremath{\stackrel{#1}{\rightarrow}}}

\newcommand{\until}[1]{\ensuremath{\mathbf{U}_{#1}}}
\newcommand{\nxt}[1]{\ensuremath{\mathbf{X}_{#1}}}
\newcommand{\always}[1]{\ensuremath{\mathbf{G}_{#1}}}
\newcommand{\eventually}[1]{\ensuremath{\mathbf{F}_{#1}}}
\newcommand{\true}{\mathtt{true}}
\newcommand{\false}{\mathtt{false}}

\usepackage{tikz}

\usepackage{pdflscape}

\usetikzlibrary{arrows,positioning,automata}

\renewcommand{\phi}{\varphi}

\newcommand{\cond}{\mathrel{|}}

\begin{document}

\newcommand{\tuple}[1]{\langle #1\rangle}
\newcommand{\sat}{\models}
\newcommand{\pr}{{\it P}}
\newcommand{\Peek}{{\it Peek}}
\newcommand{\comment}[1]{}

\newcommand{\fullOnly}[1]{}

\newcommand{\confOnly}[1]{#1}

\newcommand{\gap}{{\it gap}}
\newcommand{\pkvar}{Q}
\newcommand{\pk}{q}




\title{\bf Learning Temporal Logical Properties Discriminating ECG models of Cardiac Arrhytmias\thanks{Work partially supported by the EU-FET project QUANTICOL (nr. 600708) and
by FRA-UniTS. G.S. acknowledges support from the European Research Council under grant MLCS 306999.} \vspace*{-1ex}}

\author{
  Ezio Bartocci\inst{1}, Luca Bortolussi\inst{2,3}, Guido Sanguinetti\inst{4,5}\vspace{-1ex}
}

\institute{
Faculty of Informatics, Vienna University of Technology, Austria
\and
DMG, University of Trieste, Italy
\and
CNR/ISTI, Pisa, Italy
\and
School of Informatics, University of Edinburgh, UK
\and
SynthSys, Centre for Synthetic and Systems Biology, University of Edinburgh, UK
}

\maketitle

\begin{abstract}
We present a novel approach to learn the formulae characterising the 
emergent behaviour of a dynamical system from system observations. 
At a high level, the approach starts by devising a statistical dynamical model 
of the system which optimally fits the observations. We then propose general 
optimisation strategies for selecting high support formulae (under the learnt 
model of the system) either within a discrete set of formulae of bounded 
complexity, or a parametric family of formulae. We illustrate and apply the 
methodology on an in-depth case study of characterising cardiac malfunction 
from electro-cardiogram data, where our approach enables us to quantitatively 
determine the diagnostic power of a formula in discriminating between different 
cardiac conditions. 

\end{abstract}


\vspace*{-1ex}
\section{Introduction}
\label{sec:intro}
\vspace*{-1ex}

Dynamical systems are among the most widely used modelling frameworks, with important applications in all domains of science and engineering. Much of the attraction of dynamical systems modelling lies in the availability of effective simulation tools, enabling predictive modelling, and in the possibility of encoding complex behaviours through the interaction of multiple, simple components. This leads naturally to the notion of {\it emergent properties}, i.e. properties of the system trajectories which are a non-trivial consequence of the local interaction rules of the system components. Emergent properties of deterministic dynamical systems can often be easily verified through simulations. Quantitatively identifying the emergent properties of a stochastic system, instead, is a much harder problem.

In the simplest scenario, one assumes that a mathematical model of the system of interest is already available (e.g. as a continuous time Markov chain, or a stochastic differential equation), generally thanks to the availability of domain expertise. The problem is then somewhat the converse to the {\it model checking} problem~\cite{Clarke1982,Queille1982}: given a model, identifying which properties (out of a finite set) are verified above a certain probability threshold can be done by repeatedly checking the properties. In this contribution, instead, we consider the problem of identifying emergent properties of a system directly from observations of the state of the system at a finite number of time points. Specifically, the problem we consider is the following: given a formalisation of emergent properties as logical statements (in a suitable logic, see below) and a parametric family of formulae, identify which formula (formulae) within this family best describe the emergent behaviour of the system.

This problem, although clearly of considerable practical relevance, has received relatively little attention in the literature. We are aware of two lines of research in this direction: in early work by~\cite{Calzone2006}, a greedy algorithm was employed to identify formulae with high support directly from data, with the ultimate aim of unravelling the logical structure underpinning observed dynamics in systems biology. More recently, Asarin et al. in~\cite{Asarin2012} proposed a geometric construction to identify the formula (within a specified parametric family) which fitted observations best. In both cases, the methods work directly with the raw data, and are hence potentially vulnerable to noise in the data. Furthermore, both sets of authors remark that the identifiability of formulae is severely limited by the quantity of data available, which hampers the applicability of the methods in many practical circumstances.

Here, we aim to address both identifiability and robustness problems by taking an alternative, statistical approach. Rather than attempting directly to learn formulae from the data, we coarse-grain the data by fitting to it a statistical model which provides a compact representation of the dynamics of the system. In our case study, the statistical model will be a Hidden Markov Model (HMM): this enables us to effectively reduce the complexity of the system in a noise-aware manner, while enabling us to deploy a range of statistical tools to select the best fitting model. Once a suitable model is selected, the satisfaction probability of a formula can be evaluated quantitatively (using a model checking tool), enabling rational selection within a finite family of formulae. We can also consider infinite parametric families of formulae, as in the case of \cite{Asarin2012}; in this case we use optimisation techniques to optimise (functions of) the satisfaction probability of the formula as the parameter(s) of the formula vary. We use a recently proposed, provably convergent algorithm from the machine learning literature~\cite{gpucb}, which has previously been successfully applied in the related problem of system design by~\cite{lucaQEST13}. We illustrate our approach on a detailed case study of characterising cardiac malfunction from electro-cardiogram (ECG) data from patients with different arrhythmic cardiac conditions. Here, sections of ECG recordings are given high-level annotations defining the specific type of arrhythmia. We train different HMM models for different annotations, and then learn optimal {\it diagnostic formulae} which enable to best discriminate between different types of arrhythmia.

The rest of the paper is organized as follow: in the next section 
we introduce the case study of our approach by recapitulating some basic facts about cardiac electrical activity and related data.  In Sec.~\ref{sec:modeling} we describe how to 
learn and select HMM models with hybrid emissions (discrete for the beat type and continuous for 
the beat duration)  of signal patterns related to 
a particular cardiac arrhythmia. In Sec.~\ref{sec:properties} we 
specify the logic we will employ to 
formalise the concept of emergent behaviour and we introduce a novel  
  methodology  to learn the logical properties discriminating 
different models. In Sec.~\ref{sec:results} we present experimental results illustrating the effectiveness of the proposed approach on the case study. We 
conclude in Sec.~\ref{sec:conclusion} by discussing the implications of our contribution, 
both from the practical and the methodological aspect.


\vspace*{-1ex}
\section{Monitoring the Heart: electrophysiology and signals}
\label{sec:background}
\vspace*{-1ex}

\begin{figure}[t]
\begin{center}
\includegraphics[width=0.95\textwidth]{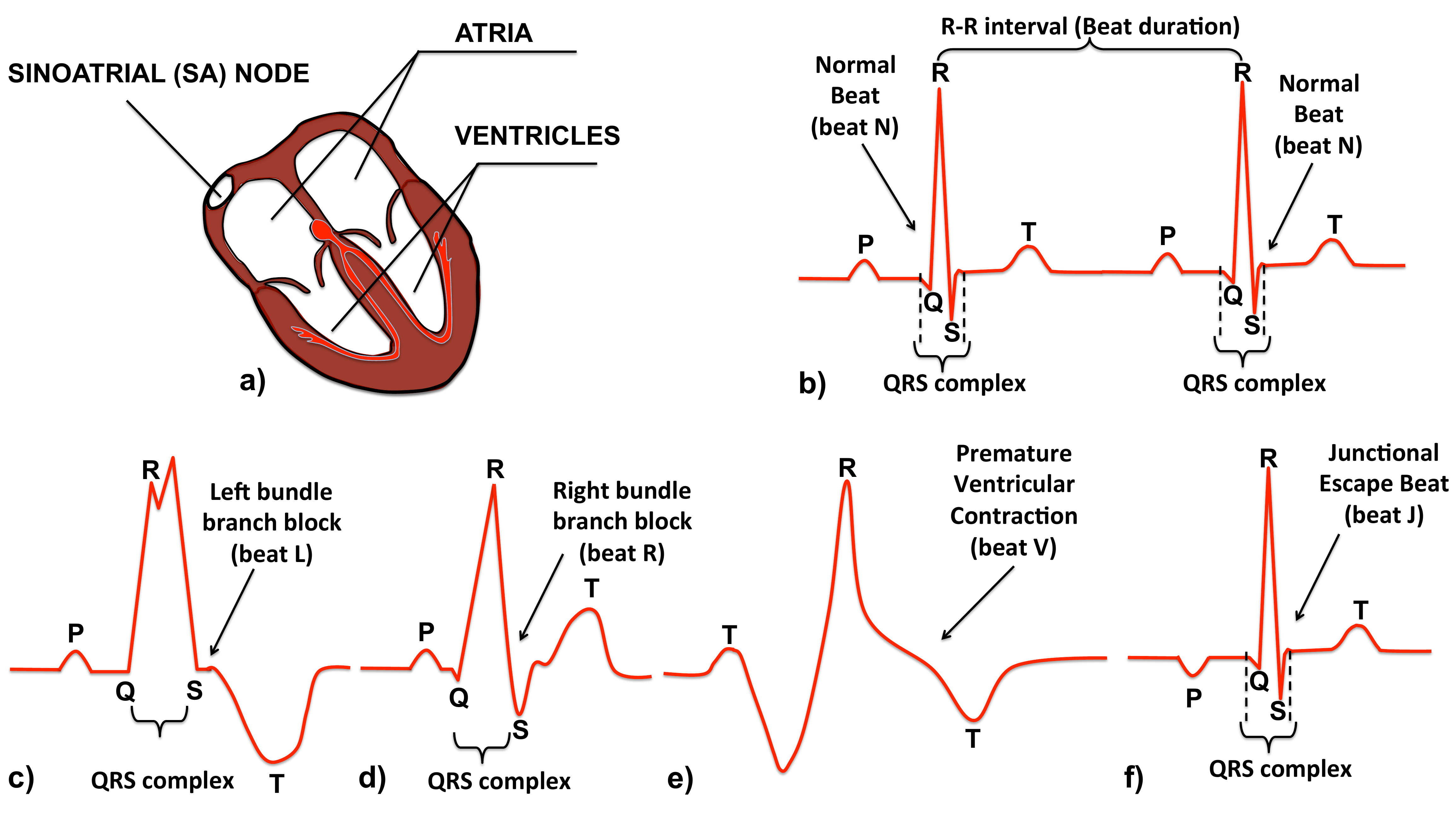}
\caption{a) Schematic representation of the heart anatomy; b) ECG pattern for two normal beats; c-f) ECG patterns for irregular beats.}
\label{figure1}
\end{center}
\vspace*{-3ex}
\end{figure}

\paragraph{\textbf{The heart in a nutshell -}}  

The heart is a muscular
pump consisting of four chambers  
(Figure~\ref{figure1} a): two upper chambers
called atria and two lower chambers  called ventricles.
The rhythmic, pump-like  function of the heart is driven by 
muscle contractions, which are in turn triggered by cell-generated 
electrical signals.
The heart consists of a network of several billions
of communicating cells, the myocytes, arranged in various
sheets and fibers, and communicating with each other through 
diffusion. The activity of each cell is regulated by a set of ionic channels, 
synchronizing each other with the membraneÕs difference of 
potential, called the action potential. 

During normal cardiac rhythm, an impulse is spontaneously generated in
the sinoatrial (SA) node 
  in the right atrium of the heart. The signal then travels  
through the atrioventricular  (AV) node that connects electrically 
the atrial and the ventricular chambers. The AV node slows down
the conduction of the pacemaker impulse, 
and this delay ensures that the all the blood in the atria passes 
into the ventricles before these contract. 
Then, the Purkinje fibers, located in the inner ventricular 
walls of the heart, synchronously  
contract the ventricles in order to pump out the blood from 
the ventricles into the circulatory system of the body.

Abnormalities in this process give rise to cardiac arrhythmias. Arrhythmias can occur both in the atria
and in the ventricles  and some of them are life-threatening 
and may result in cardiac arrest leading 
to a sudden cardiac death~\cite{adabag2010}.
Arrhythmias are a major cause of morbidity and mortality worldwide, and developing
new automated tools to assist diagnosis of arrhythmias is a challenging application domain for signal processing and computer science.

\paragraph{\textbf{The electrocardiogram (ECG) -}}  

The most common, non-invasive  diagnostic tool to  monitor the heart's electrophysiological function is the electrocardiogram (ECG).
An ECG machine is able to record the electrical activity of the heart
 through a set of electrodes (called ECG leads) placed by the 
 physician on the chest wall and limbs of the patient. 

As Figure~\ref{figure1} b) illustrates, in a healthy patient  
the ECG signal consists of three main consecutive waves: 
 \textbf{the P wave} corresponding to the depolarization  
and the consequent contraction of the atria, 
\textbf{the QRS complex} representing the rapid 
depolarization and contraction of the ventricles and
 \textbf{the T wave} identifying the recovery 
or depolarization of the ventricles. 

\paragraph{\textbf{Heartbeats -}} ECG signals are interpreted by physicians 
through a hierarchy of annotations. The fundamental unit in the ECG is the heartbeat (or, simply, beat) defined as the interval between two consecutive \textbf{R peaks}.
The beats are annotated using a symbol characterizing 
the type of beat observed (some of them shown in Figure~\ref{figure1} b-f): 

\begin{itemize}
\item  \textbf{Symbol N - Normal Beat} shown in  Figure~\ref{figure1} b).
\item \textbf{Symbol V - Premature Ventricular Contraction (PVC)}, shown
in Figure~\ref{figure1} e), is characterized by a premature wider QRS
complex, not preceded by a P wave and followed by an usually large T wave 
with an opposite concavity than in the normal beat. Physiologically, PVC beats occur earlier than expected because originating in the ventricles 
and not in the sinoatrial node.
\item \textbf{Symbol L/R - Left /Right Bundle branch block}  illustrated 
in Figure~\ref{figure1} c-d),  is an abnormal beat where one ventricle 
is delayed and contracts later than the other. 
\item \textbf{Symbol j - Junctional escape beat} in Figure~\ref{figure1} f),  is a delayed beat generated in the
atrioventricular junction. It occurs when the SA node is not working  
properly to compensate the lack of the pacemaking activity. 
\end{itemize}
Beats are usually machine annotated through pattern recognition algorithms such as support vector machines. In this work, we will use directly an annotated version of the ECG signals as a sequence of beat symbols with associated beat durations.

\begin{figure}[t]
\begin{center}
\includegraphics[width=0.95\textwidth]{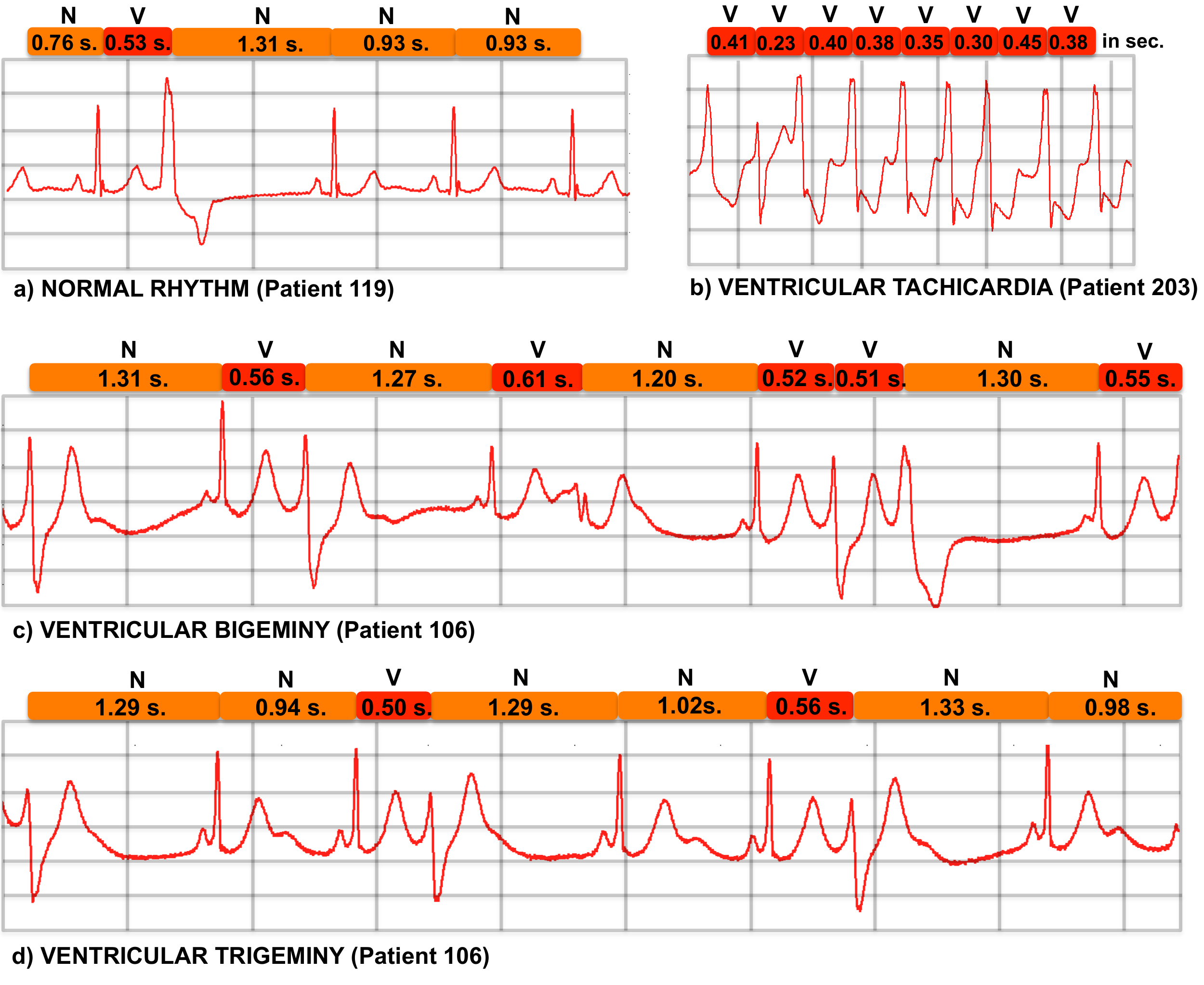}
\caption{Some ECG patterns:  a) normal sinus rhythm; 
b) ventricular tachycardia, b) ventricular bigeminy, c) ventricular trigeminy. On the top of each 
signal is reported the annotation for each beat and its duration in seconds, while on the bottom 
is reproduced the electrical signal. The ECG data was obtained from the MIT-BIH Arrhythmia Database~\cite{moody2001}}
\label{figure2}
\end{center}
\vspace*{-5ex}
\end{figure}

\paragraph{\textbf{Heart rhythms -}} A higher level annotation of ECG data is given by the {\it rhythms}, sequences of beats exhibiting a coherent pattern. Figure~\ref{figure2} a) shows an example of
an ECG pattern for a normal sinus rhythm where the initial impulse 
is generated in the SA node. Even in this case some abnormal heartbeats (such as a PVC) can sporadically occur without medical significance. In this paper we are interested in 
learning temporal logic formulas that can better discriminate potentially dangerous irregular rhythms from the normal
sinus rhythm. In particular, we focus 
on the following cardiac arrhythmias shown in Figure~\ref{figure2} b-d):

\begin{itemize}
\item \textbf{Ventricular Tachycardia} is a fast heart rhythm originating  
in one of the ventricles. 
\item \textbf{Ventricular Bigeminy} is a cardiac arrhythmia in which generally 
premature ventricular contractions alternate with normal beats. 
Figure~\ref{figure2} shows an example of ECG pattern of such arrhythmia. 
This example shows also the possibility (very rare) to have two PVCs beats 
before a normal one will occur again.
\item \textbf{Ventricular Trigeminy} is an aberrant heart rhythm consisting of
 the repetitive sequence of one premature ventricular contraction 
 followed by two normal beats.

\end{itemize}


\vspace*{-1ex}
\section{Learning models of heart rythms}
\label{sec:modeling}
\vspace*{-2ex}

%

As described in the previous section, an ECG trace is usually interpreted by physicians through a hierarchy of annotations: extended sections of the ECG trace of a patient are associated with specific rhythms. Each of these sections consists of many individual beats: each beat is in itself summarised through an annotation symbol (N, V, etc) and the time the beat lasted. Expert physicians usually provide the high level annotation of the rhythm by recognising patterns of low level symbols. Notice however that there can be still substantial symbol heterogeneity within a rythm: for example, sections annotated as normal rhythm in many patients contain many irregular beats.

Our aim is to devise a characterisation of distinct rhythms by finding logical statements that optimally discriminate between rhythms. In order to quantify the discriminative power of a formula, we use a statistical approach and determine an optimal model of the rhythm within a specific family of models, the Hidden Markov Models (HMMs). Each rhythm is assigned an HMM, which is trained on all sections of ECG recordings in a patient which are labelled with the same rhythm. In this section, we briefly review HMMs, how they can be learnt and how an optimal model structure can be selected.

\paragraph{\textbf{Hidden Markov Models (HMMs) - }}
HMMs~\cite{rabiner89hmm-tutorial} are a popular class of statistical models with applications in almost all branches of computer science, from speech recognition to natural language processing to bioinformatics. Much of their appeal lies in the ability to model complex, non-Markovian behaviour in a sequence of observations through the introduction of a sequence of unobserved, {\it latent} variables which form a Markov chain.

\begin{figure}[t]
\begin{center}
\includegraphics[width=\textwidth]{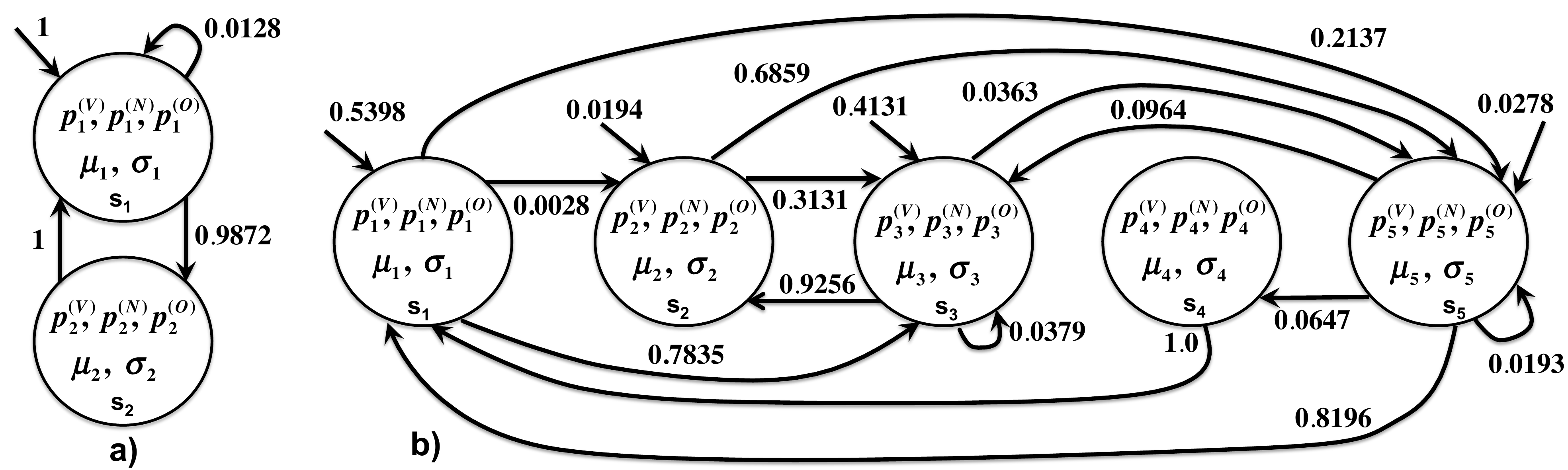}
\caption{HMM models with hybrid emission distributions for ventricular bigeminy a) and normal rhythm b). $p^{(V)}$, $p^{(N)}$, $p^{(O)}$, 
are the emission probabilities for the beats V, N, or others. 
$\mu$ and $\sigma$ are the mean and the standard deviation for the Gaussian distribution for the beat duration.}

\label{figure3}
\end{center}
\vspace*{-5ex}
\end{figure}

Formally, an HMM is a tuple
$H=\tuple{S, A, O, B, \pi}$ containing a set $S$ of states, a
transition probability distribution $A$, a set $O$ of observation
symbols, an observation probability
distribution $B$,
and an initial state distribution $\pi$ (shown schematically in Figure \ref{figure3}).  The states 
are assumed to form a finite set and are numbered, so $S$ can be written as
$S=\{s_1,s_2,\ldots,s_{N_s}\}$  where
$N_s$ is the number of states. The states are assumed to form a discrete time Markov chain; the transition probability
 $A$  of the Markov chain is an $N_s\times N_s$ matrix indexed by states in
both dimensions, such that $A_{i,j} = Pr(\mbox{state is } s_j \mbox{
  at time } t+1 \cond \mbox{ state is } s_i \mbox{ at time } t)$.  
  
The observations of an HMM are traditionally called emissions; almost all HMMs in literature are either assumed to have discrete emissions (i.e., the set $O$ is a finite set, e.g. indexing amino-acids in a bioinformatics application), or continuous emissions with observation distribution $B$ given by a mixture of Gaussians (this is the HMM-GMM model frequently used in speech recognition). In our application, the data consists of sequences of pairs of discrete symbols (the type of beat) and real-valued variables (the time the beat lasted); we therefore construct an HMM with hybrid emission probabilities given by the product of a discrete distribution (accounting for the symbol) and a normal distribution for the time. 

Three remarks are in order about this architecture: first of all, there is an implicit assumption that the duration of a beat is independent of the associated symbol, conditioned on the latent state (notice that the two are not independent marginally, i.e. when we average away the state). This assumption is difficult to verify, and we make it primarily for computational ease (assuming correlated symbols and times would lead to a proliferation of unknown parameters). Secondly, the normality assumption for the duration of the beats is another approximation: obviously, the times are positive numbers while the Gaussian distribution has support over the reals. An initial inspection of the data however showed that times are usually sufficiently far from zero, so that a normal assumption will introduce a negligible error. Finally, the states in our model do not have a physical interpretation, rather they provide a convenient device to summarize the statistical behaviour of the observed data. This is in contrast with the use of HMMs in other applications, e.g. speech recognition, where the states often represent phonemes.

\paragraph{\textbf{Inference, learning and structure learning for HMMs - }}
Having described the basic framework we work in, we are immediately faced with three tasks: how do we infer the sequence of latent states underpinning a sequence of observations? How do we determine the parameters of the HMM (initial probability, transition and emission probabilities)? Most importantly, how do we select the best fitting structure of the model (number of latent states, special structures in the transition/ emission probabilities)?

For reasons of space, we only give here a high level explanation of the statistical estimation of HMMs; we refer the interested reader to the classic review of Rabiner~\cite{rabiner89hmm-tutorial} for further details. The first problem, state inference, can be solved by exploiting the Markovian structure of the latent process, which leads to an elegant dynamic programming algorithm to solve the inference problem, the {\it forward-backward algorithm}. 

The forward-backward algorithm assumes that the parameters of the model are known: learning a (locally) optimal value for these parameters is the second of the three learning tasks delineated above. Maximum likelihood (ML) estimation provides a statistically consistent way to learn parameters (which will provably converge to the true parameter value as the amount of data increases). However, ML estimation in HMMs is complicated by the presence of the unobserved sequence of latent variables. It turns out that an iterative algorithm can be used to find a local maximum of the likelihood: given an initial estimate of the model parameters, a forward-backward pass enables state inference. The inferred state (posterior) distribution is then used to compute a lower bound on the likelihood function, which can then be easily maximised to yield an updated value of the parameters. This algorithm is called the {\it Baum-Welch algorithm} in the HMM literature~\cite{rabiner89hmm-tutorial}, and is a special case of the more general Expectation-Maximisation (EM) algorithm \cite{Dempster:maximum77}.

Finally, one has to determine which structure of HMM best fits the data. This is a discrete optimisation problem which is generally impossible to solve. Nevertheless, one can still discriminate among a finite number of models. Naively, one may be tempted to use the value of the likelihood at the maximum (computed using Baum-Welch) to select the model which best fits the data. However, it is easily seen that models with more parameters will always be preferred with this procedure, leading to the phenomenon of {\it overfitting}. One therefore needs to penalise the complexity of the model. There exist several information criteria which combine the maximum value of the likelihood with a penalty on the number of parameters. Here we use the {\it Akaike Information Criterion (AIC)} \cite{Akaike:new74}, which penalises the likelihood by subtracting a term containing the logarithm of the number of parameters. Explicitly, the AIC score is defined as 
\[AIC(i)=2k_i-2\log L_i,\]
where $k_i$ is the number of parameters of model $i$, and $L_i$ is the optimised value of the likelihood (learnt using Baum-Welch in this case). This simple score can be shown to approach asymptotically, in the large sample limit, the information lost by using model $i$ as a proxy for the (unknown) data generating process. Therefore minimisation of the AIC score across a finite number of models is often used as a criterion for model selection.
Figure \ref{figure3} shows the end product of this procedure for bigeminy (a) and normal rhythms (b). As we can see, bigeminy is essentially captured by a model with only two states, rapidly transiting between the two states and emitting either N or V symbols preferentially in each state. The normal rhythm HMM instead has a more complex architecture, reflecting the fact that transient abnormal beats can also be present in overall normal rhythm.

\vspace*{-1ex}
\section{Learning to discriminate cardiac rhythms}
\label{sec:properties}
\vspace*{-1ex}

In this section, we introduce the main methodological steps we use. We start by briefly reviewing Metric Interval Temporal Logic \cite{Alur1996}, the logical framework we work in, and formally introduce the concept of discriminative formulae, i.e. the objective function of our learning procedure. We then describe how we construct simple template formulae by combining atomic formulae on the observation symbols and evaluate their discriminatory power. Finally, we combine highly discriminative template formulae in a larger parametric formula (stating that these templates are repeated for a certain time within a rhythm) and optimise the parametric formula using a provably convergent optimisation algorithm from the reinforcement learning literature \cite{gpucb}.

\vspace*{-1ex}

\paragraph{\textbf{Metric Interval Temporal Logic - }}
Temporal logic~\cite{Pnueli1977} provides a very elegant framework to specify 
in a compact and formal way an emergent behaviour in terms of {\it time-dependent} events.
Among the myriads of temporal logic extensions available, Metric Interval Temporal 
Logic~\cite{Alur1996} (MITL) is very suitable to characterise patterns of heartbeats, as it describes properties of a discrete signal evolving in continuous time. The syntax of MITL is as follows:

\begin{definition} [STL syntax] The syntax of MITL is given by
$$\varphi := \true \:|\:  q \:|\:  \neg \varphi \:|\: \varphi_{1} \wedge \varphi_{2} \:|\: \nxt{ [a,b] } \varphi  \:|\:  \varphi_{1} \: \until{[a,b]} \: \varphi_{2},$$
where $\true$ is a true formula, $q$ is an atomic proposition which is either true or false in each state $S$, conjunction and negation are the standard boolean
connectives, $[a,b]$ is a dense-time interval with $a<b$, $\nxt{[a,b]}$ is the \emph{next}  operator and  $\until{[a,b]}$ is the {\it until} operator. 
\end{definition}
The (bounded) until operator $\varphi_{1} \: \until{[a,b]} \: \varphi_{2}$ requires $\varphi_{1}$ 
to hold from now until, in a time between $a$ and $b$ time units, $\varphi_{2}$ becomes true, while the (bounded) next operator $\nxt{[a,b]}\varphi$ requires $\phi$ to hold in the next state, to be reached between  $a$ and $b$ units of time. The {\it eventually} operator  $\eventually{[a,b]}$ and  the {\it always } operator $\always{[a,b]}$ can be defined as usual: $\eventually{[a,b]}  \varphi := \top \until{[a,b)} \varphi$, $\always{[a,b]} \varphi := \neg \eventually{[a,b]} \neg \varphi.$

More precisely,  MITL has two main semantics, according to the nature of the paths in which it is interpreted \cite{MITLsurvey}. In the \emph{pointwise semantics}, paths are sequences of states from the (discrete) state space $S$, and time delays between them, usually represented as $\sigma=s_0\lts{t_0}s_1\lts{t_1}s_2\lts{t_2}\ldots$. In this semantics, defining $T_i = \sum_{j=0}^{i-1} t_i$, the time at which we enter the $i$-th state, and $\sigma^i$ the  suffix of $\sigma$ starting at position $i$, we have that $\sigma, t \models \phi_{1}\until{[a,b]}\phi_2$ if and only if $\exists j \geq 0$ such $\sigma^j,t + T_j \models \phi_2$, $T_j \in [a,b]$, and for each $0\leq i < j$, $\sigma^i,t + T_i \models \phi_1$. The definition is similar for the next operator: $\sigma, t \models \nxt{[a,b]}\phi$ if and only if $\sigma^1,t + T_1 \models \phi$ and $T_1\in[a,b]$.\\
The alternative  \emph{continuous semantics} treats time as a proper continuous entity and considers boolean signals, which are are functions of time to $\{\true,\false \}$, as atomic propositions. See \cite{Maler2004} for further details on the logic and the monitoring algorithm. 

We will focus on the pointwise semantics, as it is the standard interpretation of MITL for (Continuous Time) Markov Chains (see for instance \cite{Zuliani2009}), 
where the truth of  atomic propositions  depends only on the discrete observables (as opposed to the latent states). For this reason, and for the presence of continuous time, our approach to specify properties of HMM is different from that of \cite{Zhang2005}. Furthermore, 
the non-exponential nature of random times in hybrid HMM models (and the use of MITL to specify properties) makes the use of numerical algorithms very difficult. We are not aware of methods to model check such a class of stochastic systems, although it may be possible to recast our problem in a form suitable for the stochastic class approach of \cite{Vicario2012}. Hence, we resort to Monte Carlo methods, applying statistical model checking (SMC) to estimate the probability of a MITL formula.
Generally, the models learned will have a small set of latent states, hence simulation turns to be very fast.

\vspace{-1ex}
\paragraph{\textbf{Statistical Model Checking - }}
SMC is a Monte Carlo approach to estimate the satisfaction probability of a (MITL) formula against a stochastic model, or to test if this probability is above or below a given threshold. Roughly speaking, it consists of generating many samples of the stochastic model, using a simulation routine, and monitoring the value of the MITL formula against the so obtained traces. This procedure generates samples from a Bernoulli random variable with unknown probability $p=P(\phi)$, which can be estimated using standard techniques. In this work, we use a Bayesian estimation scheme \cite{Zuliani2009}, imposing a non-informative Beta prior on $p$ (with both parameters equal to one). This corresponds to a regularisation of the estimate, starting to count the number of successes and failures from one rather than from zero. This  avoids problems when the true probability is very close to zero or to one. 

\vspace{-1ex}

\paragraph{\textbf{Discrimination score - }} 
Consider a fixed MITL formula $\phi$ and two HMM models $M_1$ and $M_2$, learned from the same patient but for two different rhythms (e.g. bigeminy versus normal). We consider $\phi$ a good  formula to discriminate between  $M_1$ from $M_2$ if it is satisfied {with high probability by trajectories of $M_1$ \textbf{and} satisfied with low probability by trajectories of $M_2$. The rationale for this definition is practical: recall that our formulae are defined on observables. Given a new sequence with unknown annotation, checking the formula on this sequence would allow high confidence labelling of the sequence.} Calling $P(\phi|M_i)$ the probability of $\phi$ in the model $M_i$, we can quantify this discrepancy by the \emph{log-odd ratio} 
\begin{equation} \label{discrimination} 
R(\phi,M_1,M_2) = \log\left(\frac{P(\phi | M_1)}{P(\phi | M_2)}\right).
\end{equation}
This quantity varies between $-\infty$ and $+\infty$, and is positive and large if $\phi$ supports $M_1$ against $M_2$, negative and large if $\phi$ is more likely in $M_2$ than in $M_1$, and zero if it has the same probability in both models. In particular, it equals  $+\infty$ when $P(\phi | M_2)$ is zero and 
$-\infty$ when $P(\phi | M_1)$ is zero. However, as we will estimate $R$ by SMC using the Bayesian regularisation, we are guaranteed that its value will always be finite, as $0<P(\phi | M_i)<1$.

\vspace{-1.6ex}
%
%
\paragraph{\textbf{Learning Discriminative Formulae - }}
Identifying formulae with high discriminative power can be cast as a mixed integer/ continuous optimisation problem. One needs in fact  to identify the structure (or template) of the formula (out of a discrete set of possible combinations of atomic propositions) and the set of continuous parameters entering the formula which maximise the discriminative power (in the heart example, these continuous parameters represent bounds on the durations of the beats).
 
Finding good formula templates is a difficult task, and requires insights on the system under examination.  In this paper we do not tackle the problem in its full generality, but we set up a greedy search scheme which can easily incorporate some basic knowledge of the domain at hand. On the other hand, identifying the parameters of a formula of a fixed structure which maximise its discriminative power can be done elegantly and effectively borrowing ideas from {\it reinforcement learning}.

More specifically, we assume that we have a MITL formula $\phi_{\theta}$ which depends on some continuous parameters $\theta$. We aim to maximise its discriminative power $R(\theta) = R(\phi_{\theta},M_1,M_2)$ defined in equation \eqref{discrimination}. Naturally, this quantity is an intractable function of the formula parameters; its value at a finite set of parameters can be noisily estimated using an SMC procedure. The problem is therefore to identify the maximum of an intractable function with as few (approximate) function evaluations as possible. This problem is closely related to the central problem of reinforcement learning of determining the optimal policy of an agent with as little exploration of the space of actions as possible. We therefore adopt a provably convergent stochastic optimisation algorithm, the GP-UCB algorithm \cite{gpucb}, to solve the problem of continuous optimisation of formula parameters. Intuitively, the algorithm interpolates the noisy observations using a stochastic process (a procedure called emulation in statistics) and uses the uncertainty in this fit to determine regions where the true maximum can lie. This algorithm has already been used in a formal modelling scenario in \cite{lucaQEST13}.

We now turn to describe the general optimisation procedure. We assume to have a fixed set of \emph{basic} template formulae $\mathcal{T}$ (typically derived from domain expertise). 
Fix two models $M_1$ and $M_2$, and assume we want to discriminate the former from the latter.
First, we search exhaustively in $\mathcal{T}$ by optimising the continuous parameters of each $\phi\in\mathcal{T}$, and thus computing its best score (i.e. the maximum log-odd ratio). Then, we rank the formulae in $\mathcal{T}$ and select the subset of higher score,  provided this score is  larger than zero, and the satisfaction probability for $M_1$ is sufficiently high (in the context of this paper, greater than 0.2).  If this yields a non-empty set of formulae $\mathcal{T}_{best}$, we proceed to the second phase, otherwise we enlarge the set $\mathcal{T}$, and restart the procedure. 
%
%

In the second phase, we take the  formulae in $\mathcal{T}_{best}$ and combine them using some predefined combination rules (for instance, boolean combinations), and run again  the continuous optimisation on the parameters, ranking  the formulae and selecting those with highest log-odds score. 
As the set $\mathcal{T}_{best}$ is expected to be small, we will be searching exhaustively a reasonably small set of formulae. At this stage, we expect this greedy optimisation to have found some good discriminating formula, i.e. a formula with a good log-odd ratio score, having high probability in $M_1$ and low probability in $M_2$. If not, we can  combine together the best formulae of this second round, possibly with another set of combinators, or reconsider the choice of the basic templates $\mathcal{T}$.

\vspace{-1ex}

\paragraph{\textbf{Template formulae for the heart - }}  The characteristic hallmark of an abnormal heart rhythm is in the presence of a certain pattern of  heartbeats. These could be consecutive or non-consecutive sequences of specific beats. We are precisely looking for these patterns, with additional information on the duration of beats. Hence, the basic template formulae need to describe short patterns of consecutive sequences of beats, while the greedy discrete search procedure needs to combine them to provide more complex patterns.
More specifically, we consider basic template formulae of the form 
\[ \phi := \eventually{}\always{\leq T} \phi_{o_1\ldots o_k};\] 
where $\phi_{o_1\ldots o_k}$ is recursively defined by
$ \phi_{o_1\ldots o_k} := q_{o_1} \wedge \nxt{[a_{o_1},b_{o_1} ]} \phi_{o_2\ldots o_k}$,  and $\phi_{\emptyset} := \true.$
The atomic proposition $q_{o_j}$ is true when we observe the symbol $o_j$. Hence,  the formula $\phi_{o_1\ldots o_k}$  identifies the sequence of observations (heartbeats) $o_1\ldots o_k$, further imposing a time constraint on the duration of each beat: the duration of  $o_j$ in constrained between $a_{o_j}$ and $b_{o_j}$. The formula $\eventually{}\always{\leq T} \phi_{o_1\ldots o_k}$ additionally imposes that this sequence can be found somewhere in the signal under observation, and has to last for at least $T$ units of time.
When running the GP-UCB algorithm, we optimise jointly all the temporal constraints: the duration of symbols $o_j$ and the parameter $T$ of the always operator. We considered formulae for a small value of $k$, between 2 and 4. 

As far as the combination phase is concerned, we simply take the disjunction of a subset $O\subset \mathcal{T}_{best}$ of the best formulae $\phi_{o_1\ldots o_k}$ found in the first phase of the search. Essentially, we are considering formulae of the form
\[ \phi_O := \eventually{}\always{\leq T} \bigvee_{o_1\ldots o_k\in O} \phi_{o_1\ldots o_k},\]
where $O$ ranges among all subsets of size 2 or more of $\mathcal{T}_{best}$.
This way of combining formulae aims at identifying a subset of patterns that is strongly expressed in an arrhythmia.

\vspace*{-0.5ex}
\section{Results}
\label{sec:results}
\vspace*{-0.5ex}


\noindent\textbf{\emph{Experimental setup - }} We present here initial results on annotated ECG data from the MIT-BIH Arrhythmia Database~\cite{moody2001}. We restricted our attention to a subset of possible rhythms which were more prevalent in the data: bigeminy, trigeminy, ventricular tachycardia and  the normal rhythm. These signals are predominantly composed of  \textbf{ V} and \textbf{ N} symbols, often with a similar frequency, hence discrimination is more challenging. Due to space restrictions, we present results on a single patient (patient 233);  other patients yielded similar results. Code to recreate the experimental results is available from the authors for academic use. The experimental procedure can be summarised as follows 
\begin{itemize}
\item For each rhythm, we learn HMM models with 2 to 6 states, and select the one with best AIC score. We learn the model simultaneously on all segments annotated as a certain rhythm (e.g. bigeminy). 
\item For each pair of abnormal/ normal rhythm, we learn template formulae starting from the basic set of formulae $\mathcal{T}_2$, corresponding to possible patterns of length 2 of symbols V and  N: $\mathcal{T}_2 = \{\eventually{}\always{\leq T} \phi_{NN}, \eventually{}\always{\leq T} \phi_{NV}, \eventually{}\always{\leq T} \phi_{VN}, \eventually{}\always{\leq T} \phi_{VV}\}$,  and optimise the continuous parameters ($T,b_N,b_V$) to give maximum discriminative power \footnote{We seach in the following space: maximal duration of symbols is constrained between 0 and 2.5 seconds, while the lower bound was set to zero. The total duration $T$ varies between 0 and an upper bound  depending on the signal, equal to 4 for bigeminy, 7 for trigeminy, 2 for tachycardia. We generate signals of fifteen seconds. The choice of bounds for $T$ is  consistent with the duration of raw signals in the training set.}.
\item If after the optimisation phase no highly supported formula was found, i.e. a formula with high log-odd ratio of abnormal versus normal signal and high satisfaction probability, we rerun the procedure increasing the pattern length of one (hence, first for $\mathcal{T}_3$, then  $\mathcal{T}_4$, and so on).
\item We then selected the most supported formulae of $\mathcal{T}_k$
to further combine them, as discussed in the previous section. We run the continuous optimisation also for these formulae, and chose the  ones having both high log-odd ratio and satisfaction probability for the abnormal rhythm.
\end{itemize}
We now present results on discrimination of the three abnormal rhythms in more detail.

\

\noindent\textbf{\emph{Bigeminy - }} 
%
%
Learning formula templates for the discrimination of bigeminy against normal heart behaviour proceeded as follows: in the first optimisation run, the two formulae with highest log-odd ratio where $\eventually{}\always{\leq T} \phi_{NV}$ and $\eventually{}\always{\leq T} \phi_{VN}$, scoring more than 5, with a satisfaction probability in bigeminy of about $0.8$. 
The other two formulae, instead, have a log-odd ratio zero or less.
%
 Hence, we selected these two formulae for the second phase of the discrete search, obtaining $\eventually{}\always{\leq T} \phi_{NV} \vee \phi_{VN}$ as the only candidate for the second round. This formula clearly codes for the pattern $VN$ repeated many times (for as long as $T$ units of time). 
Running the continuous optimisation, we find a log-odd ratio of 4.08, which is lower than in the previous case, but it corresponds to a satisfaction probability of 0.9994 in the abnormal rhythm, and a probability of 0.016 in the normal one, corresponding to a sensitivity of $>99$\% and a specificity of approximately 98\%. Hence, this formula turns to have a good discriminative power, and its relatively low log-odd ratio depends on its high sensitivity to small values of the denominator. The upper bound of time $T$  is optimally set to $3.8$, close to the maximum of $4$. Upper bounds on beat duration are also close to their maximum. 
%
%
Note that the alternation of $V$ and $N$ is precisely what characterises bigeminy: our method learned the correct pattern used by physicians, and additionally quantitated the time such a pattern persists for.

%

\noindent\textbf{\emph{Trigeminy - }} To discriminate trigeminy vs normal rhythm, we proceeded analogously as for bigeminy, starting with the same set $\mathcal{T}_2$ of formulae. In this case, however, no formula of length 2 was found to have a high support in discriminating trigeminy (less than 3.5), hence we considered basic templates corresponding to patterns of length 3. The analysis in this case gave high log-odd ratio (4 or greater) to three formulae: $\eventually{}\always{\leq T} \phi_{VNN}$, $\eventually{}\always{\leq T} \phi_{NVN}$, and $\eventually{}\always{\leq T} \phi_{NNV}$, with for a small duration $T$ for all three cases. We then took all possible combination of at least two of those formula using disjunction, and found  the most discriminating  formula (log-odd ratio 7.8, satisfaction probability for trigeminy $0.9968$, and for normal signal of $0.004$) to be $\eventually{}\always{\leq T} \phi_{VNN} \vee \phi_{NVN} \vee \phi_{NNV}$, corresponding to the pattern $VNN$ repeating in time for approximatively $T=4.25$ seconds. Again in this case, the method found the hallmark pattern of trigeminyand additionally quantified its persistent behaviour. We also tested that this formula works well in discriminating trigeminy versus bigeminy (log-odd ratio of 8.5).

\noindent\textbf{\emph{Ventricular tachycardia - }} This case turned out to be the simplest one. A good disciminating formula was found already in the set $\mathcal{T}_2$, corresponding to the pattern $VV$. In particular, the continuous optimisation returned a log-odd ratio of 2.9, corresponding to a satisfaction probability in the abnormal rhythm of 0.9998 and of $0.05$ in the normal rhythm, with the global validity time $T$ set approximatively to 1.25  seconds. This corresponds to  tachycardia being characterised by a stretch of about 3 to 4 $V$ beats. 

\noindent\textbf{\emph{Discrimination on other patients - }} So far, we considered discriminative power as applied to the same patient on which the models were learnt. We now consider the much harder task of assessing whether formulae remain discriminative when also applied to other patients. We considered three other patients for each arrhythmia, and obtained an high discriminative power, as reported in Table \ref{table:patients}. 
We also tested the formulae on raw signals taken from the database\footnote{22 signals for bigeminy of length at least 4.5, 49 for trigeminy of length at least 5, and about 80  for normal rhythms. We did not treat tachycardia because there were too few signals.}, obtaining the following results, in terms of satisfaction probability: 0.954 for bigeminy versus 0.038 for normal rhythms (on the same patients); and 0.918 for trigeminy versus 0.287 for normal rhythms. The high satisfaction probability on normal rhythms for this last case is almost entirely explained by the presence, in the extended data set, of several patients with slow heart beats. The relatively low duration of the pattern learnt on the training patient, 4.25s, can be matched in the slow beat patients by patterns consisting of a single V preceded and followed by two Ns, which are very common in normal rhythm. Increasing the time bound to 7s reduced the satisfaction probability in normal rhythms to 0.014, while the satisfaction probability for trigeminy remained stable to 0.906.

%
%

\begin{table}[!t]
\begin{center}
\begin{tabular}{|cccc|}
\hline
 & Bigeminy\ \  & Trigeminy\ \  & V. Tachycardia \\
\hline
Av. log-odd ratio & 3.32 & 2.99 & 7.68 \\ 
\hline 
Av. prob. abnormal &  0.99 & 0.99 & 0.99\\
\hline
Av. prob. normal & 0.06 & 0.08 & 0.0005\\
\hline
\end{tabular}
\end{center}
\caption{Average log-odd ratio and satisfaction probability of abnormal and normal signals for the best discriminating formulae learned from patient 223, as tested on other three patients (per type of abnormality). Tested on patients: 119, 213, and 233 for bigeminy; 119, 201,208 for trigeminy; 213, 215, 233 for ventricular tachycardia.}
\vspace{-0.6cm}
\label{table:patients}
\end{table}

\vspace*{-2ex}
\section{Conclusion}
\label{sec:conclusion}
\vspace*{-1ex}
This paper proposes a novel approach to the general problem of learning the 
emergent properties of a system from observations, illustrating the approach on 
an important biomedical case study. As such, we aim to give 
contributions both from the methodological point of view and on the applicative side.

From the methodological point of view, the statistical treatment of the observations advocated in this paper 
offers several advantages over earlier approaches which utilised the raw data in an inductive 
fashion~\cite{Calzone2006,Asarin2012,grosu2009}.The availability of a statistical model 
enables a principled handling of noise in the data, potentially leading to increased robustness in 
the predictions, and naturally gives quantitative semantics for the satisfaction probabilities of formulae 
(which is not available e.g. in the approach of \cite{Calzone2006}). 
Furthermore, the statistical model enables us to leverage the full arsenal of statistical 
 machine learning methodologies, e.g. in order to select the most fitting model structure, or to incorporate 
 relevant prior information in the models. In this sense, our paper is part of a growing trend of combining 
 advanced machine learning methodologies with formal modelling 
 techniques~\cite{lucaQEST13}, with the ultimate aim of combining statistical 
 principles with formal reasoning.

From the application point of view, ECG data and other physiological signals 
have received enormous interest both in the formal modelling 
community~\cite{bartocci2009,grosu2011,chen2012,jiang2012,chen2013} and in the machine learning 
community \cite[e.g.]{Skolidis:Bayesian11,Quinn:factorial09}. Our hybrid formal-learning approach is, to our 
knowledge, novel within this application domain, and may offer advantages. For example, while it is relatively 
easy to learn classifiers on ECG data with excellent performance on a single subject, transferring the classifier 
learnt on one subject to a novel subject is very difficult, and to some extent still an open 
problem~\cite{Skolidis:case12}: our approach, based on learning high support discriminative formulae, appears 
to be more robust to this end. 

We further stress that the learning algorithm we discussed produces as most discriminating formulae precisely those corresponding to the known patterns of heartbeats that characterise the arrhythmic phenomena considered, and additionally enables a quantification of the time that these patterns persist for.  These formulae also have a good predictive power in discriminating normal and abnormal signals. However, learning a temporal logic formula is conceptually different than learning a good predictive classifier based on a stochastic model. In fact, a temporal logic formula can immediately be understood by humans, being the description of  a causal pattern between events. In this sense, learning logical properties gives insights in the phenomenon at hand: they have an intrinsic explanatory power and provide an understandable synthesis of the most relevant features of observed data. Furthermore, this logical characterisation opens up the possibility of (automatically) synthesing run-time monitor algorithms to detect dangerous cardiac rhythms at their onset.


\vspace*{-1ex}
\bibliographystyle{splncs03}
\bibliography{tacas}

\end{document}